\documentclass[11pt]{article}
\usepackage[margin=1.0in]{geometry}
\usepackage[T1]{fontenc}
\usepackage[utf8]{inputenc}
\usepackage{lmodern}
\usepackage{microtype}
\usepackage{graphicx}
\graphicspath{{figures/}}
\usepackage{amsmath,amssymb,bm}
\usepackage{booktabs,tabularx,array,longtable,multirow}
\usepackage{caption}
\usepackage{float}
\usepackage{authblk}
\usepackage{placeins}
\usepackage[numbers,sort&compress]{natbib}
\usepackage[hidelinks]{hyperref}
\usepackage{xcolor}
\usepackage{enumitem}
\setlist{nosep,leftmargin=*}
\newcommand{\RR}{\mathbb{R}}

\newcommand{\rd}{\mathrm{RD}}

\newcommand{\bdm}{\mathrm{BDM}}
\newcommand{\ctm}{\mathrm{CTM}}

\begin{document}
\title{Canonical residual physics-informed neural networks and algorithmic perturbation landscapes map cell-fate transitions in longitudinal H3K27M glioma}

\author[1,2,3,*]{Abicumaran Uthamacumaran}

\affil[1]{Department of Pathology and Laboratory Medicine, University of British Columbia, Canada}
\affil[2]{Oxford Immune Algorithmics, Reading, United Kingdom}
\affil[3]{Department of Surgical and Interventional Sciences, McGill University, Montreal, Canada}
\affil[*]{Corresponding author: \texttt{a\_utham@live.concordia.ca}}

\date{}
\maketitle


\begin{abstract}
Pediatric high-grade glioma is a dynamically heterogeneous disease in which therapeutic selection pressures, microenvironmental selection and developmental plasticity can reorganize cellular-state structure over time. Here we investigated these adaptive behaviors in glioma dynamics from single-cell ecologies by merging physics-driven artificial intelligence and complex systems theory. We developed a single-cell latent space manifold framework that combines cancer complexity and graph networks-based algorithmic information dynamics. Specifically, this study explores the intersection of canonical residual-based physics-informed neural networks (PINNs), dynamical systems modelling comparators, explainable artificial intelligence (XAI), and Block Decomposition Method (BDM) perturbation analysis to understand glioma cell fate decisions. Two independently trained coordinate-field PINNs were evaluated: a reaction--diffusion (RD) model enforcing $q_t-Dq_{xx}-R_\theta(q)=0$, and a Fokker--Planck (FP) model enforcing probability conservation together with latent advection--diffusion. Both used automatic-differentiation residuals at collocation points, explicit initial and no-flux boundary conditions, data anchors, Adam warm-up and strong-Wolfe L-BFGS refinement. The primary analyses distinguished longitudinal initial from recurrent H3K27M cells and, within the recurrent specimen set, stem/progenitor-like from differentiated/glial-like annotation-derived states. Across five fold cross-validation, RD achieved AUC $0.995\pm0.005$ and macro-F1 $0.973\pm0.016$ for initial--recurrence discrimination, and AUC $0.698\pm0.012$ and macro-F1 $0.629\pm0.014$ for recurrent-state subtype discrimination. FP achieved AUC $0.931\pm0.018$ and $0.665\pm0.012$, respectively. Cross-model XAI repeatedly prioritized \textit{PTGDS}, \textit{CHI3L1}, \textit{IGFBP7}, \textit{SPARCL1} and \textit{NRXN3} in longitudinal progression, and \textit{CHI3L1}, \textit{MAPK4}, \textit{LRP1B}, \textit{KAZN} and \textit{RBFOX1} within recurrent-state structure. BDM perturbation rankings emphasized heat-shock/stress-response and immune-associated network signatures, including \textit{HSPA1A}, \textit{HSPH1}, \textit{HSPB1}, \textit{INPP5D} and \textit{APBB1IP}, while learned latent space BDM shifts showed marked class and dimension dependence. These findings are candidate computational network biomarkers rather than experimentally established causal signatures. The study provides a reproducible bridge between physics-constrained latent dynamics, interpretable gene attribution, and data-driven algorithmic perturbation landscapes for in silico pattern generation about glioma cell-fate (developmental) plasticity and early recurrence detection (predictive forecasting).
\end{abstract}


\section{Introduction}
Complex systems science, or simply complexity theory, studies how interactions (relationships), feedback, and constraints among multi-agents or components of a network produce collective behaviors (emergent dynamics) and multi-nested patterns that cannot be reduced to or be inferred by examining the parts in isolation \citep{bossomaier2000complex,ladyman2020complex,anderson1972more}. In this sense, complexity theory is a science of patterns, relations and wholes: network architecture defines which interactions and recursive multiscale feedback loops are possible, dynamical systems define how states evolve, and multi-dimensional constraints shape the emergent repertoire of behaviors. Cancers, including aggressive pediatric high-grade gliomas (pHGGs), can be conceptualized as complex adaptive ecosystems whose phenotypic states emerge from nonlinear gene-regulatory network dynamics and attractor-like landscapes \cite{uthamacumaran2025phgg,uthamacumaran2022gbm,uthamacumaran2020complex,huang2009cancer,huang2005cellfates,aguade2024ecosystems}. Tumour ecosystems comprise adaptive interactions among diverse tumour, stromal, and immune cell populations, as well as developmental-like cellular identities such as stem and progenitor states; under treatment-selection pressures and stalled developmental programs, these interactions form an evolving ecosystem whose phenotypic states can reorganize through plasticity, selection, and feedback\citep{hanahan2022hallmarks,neftel2019integrative,filbin2018developmental}. The resulting cell fate trajectories are not merely on/off switches of differentially expressed genes. They are complex adaptive networks creating paths through a state-space containing basins, barriers, branching regions (bifurcations) and when stuck, potentially maladaptive attractors\cite{uthamacumaran2025phgg,uthamacumaran2022gbm,uthamacumaran2020complex,huang2009cancer,huang2005cellfates,aguade2024ecosystems}. As traditional science has largely focused on snapshots of these processes and reductionistic or statistically bound trajectory inference, the complex dynamics of cell fate transitions in glioma ecology remain poorly understood, and we lack an intuitive physics of how these patterns emerge or evolve.

Longitudinal single-cell profiling make these questions empirically accessible. The recent Sussman et al. pHGG atlas combines single-nucleus and spatial multiomic observations across pediatric high-grade glioma evolution and treatment response \citep{sussman2026atlas}. Such time-resolved patient data remain observational, sparse and irregular in time limiting the applicability of tools utilized in conventional time-series forecasting, such as delay embedding and regression analyses. Pseudotime and low-dimensional manifolds can organize cells as trajectory inference, but neither should be treated as chronological time, anatomical space (morphospace) or causal inference. The methodological challenge is therefore to construct predictive, causal, and interpretable latent dynamics capturing morphospace (i.e., the state space of all possible pattern formation in single-cell fate decisions) while keeping the status of the inferred coordinates explicit.

Physics-informed neural networks offer one such route. A canonical residual PINN is more than a typical neural network with a physics-motivated penalty: it represents a coordinate-dependent field and minimizes automatic-differentiation residuals of governing differential equations at collocation points, together with initial, boundary and data constraints \citep{raissi2019pinn}. Such systems allow an intuitive form of representation and manifold learning of complex cell fate geometries in decision-making processes. We adapted two open-source PINN layouts \citep{bhustaliPINNrepo,kimPINNrepo} to single-cell latent manifolds. One model represents interacting Turing morphogenesis, i.e., activator/inhibitor-like latent fields, by a learned reaction--diffusion system \citep{turing1952chemical}. The second represents probability transport by Fokker--Planck conservation coupled to latent advection--diffusion. In cellular-state space, Fokker--Planck dynamics describe how a probability density of cells evolves under deterministic drift and stochastic diffusion, capturing population flow, dispersion, and redistribution along cell fate trajectories \citep{risken1996fokker}. Neural ODEs, graph neural networks, a Schrödinger bridge and physics-structured latent regularizers were retained as broader complex dynamics comparators \citep{chen2018neuralode,debruijn1956fokkerplanck,leonard2014schrodinger}.

Neural ODEs parameterize a latent vector field, $\dot z=f_\theta(z,t)$, and obtain representations by numerical integration, providing a flexible continuous-depth alternative to fixed discrete transformations \citep{chen2018neuralode}. Residual PINNs instead constrain coordinate-dependent fields through specified differential operators and initial/boundary conditions \citep{raissi2019pinn,karniadakis2021physics}. Their comparison therefore asks whether explicit reaction, diffusion or conservation structure provides a useful inductive bias beyond a learned latent vector field to capture the collective dynamics and emergent patterns in glioma ecology/cell fate decisions. In this application, these structures are candidates or in silico hypotheses about the constructed manifold and morphogenetic processes steering glioma cell fate dynamics.

Prediction alone does not identify which molecular features organize a transition. We therefore linked latent dynamics to decoder weights, Integrated Gradients and reaction-field Jacobian projections \citep{sundararajan2017axiomatic}. We then reconstructed feature-derived gene networks and quantified node, edge and latent perturbations with the Block Decomposition Method (BDM), implemented with \texttt{pyBDM} \citep{talaga2024pybdm}. BDM is related to, but not identical with, Minimum Description Length (MDL). MDL selects a model by minimizing $L(M)+L(D\mid M)$ \citep{rissanen1978model}; Kolmogorov complexity computes the shortest program generating an object \citep{kolmogorov1965,chaitin1966}; BDM approximates algorithmic description length by decomposing a large object into small blocks for which coding-theorem estimates are available \citep{delahaye2012numerical,zenil2018decomposition}. Under perturbation, $\Delta\bdm$ measures how the estimated description length changes when a component is removed or modified. Algorithmic Information Dynamics (AID) uses such interventions to prioritize candidate generative mechanisms capable of steering or reprogramming these complex systems, but in biological applications these remain computational causal candidates until experimentally tested \citep{zenil2019calculus,zenil2023aid}. BDM perturbation functions as an in silico causal inference method by systematically perturbing nodes or edges and measuring the resulting change in estimated algorithmic complexity. Graph network elements whose removal or modification produces large $\Delta\mathrm{BDM}$ are interpreted as having stronger causal contributions to the system’s generative dynamics and reprogrammability \citep{zenil2019calculus,zenil2023aid}.

Here we ask whether canonical residual-based latent PINNs can map longitudinal and intrarecurrence cell-state structure, whether XAI can identify stable transition-associated genes across models and objectives, and whether BDM perturbation landscapes add information distinct from statistical information theoretics such as entropy or lossless compression. The emphasis is not a claim that pseudo-space obeys literal tissue physics and morphogenesis, but rather than these complex systems methods simulate their interactions and emergent behaviors. That is, differential equations and physics-driven neural networks are used as explicit structural hypotheses over a fold-derived latent manifold, producing mechanistically interpretable fields, residuals and intervention-sensitive representations.

\begin{figure}[H]
\centering\includegraphics[width=\linewidth,height=0.78\textheight,keepaspectratio]{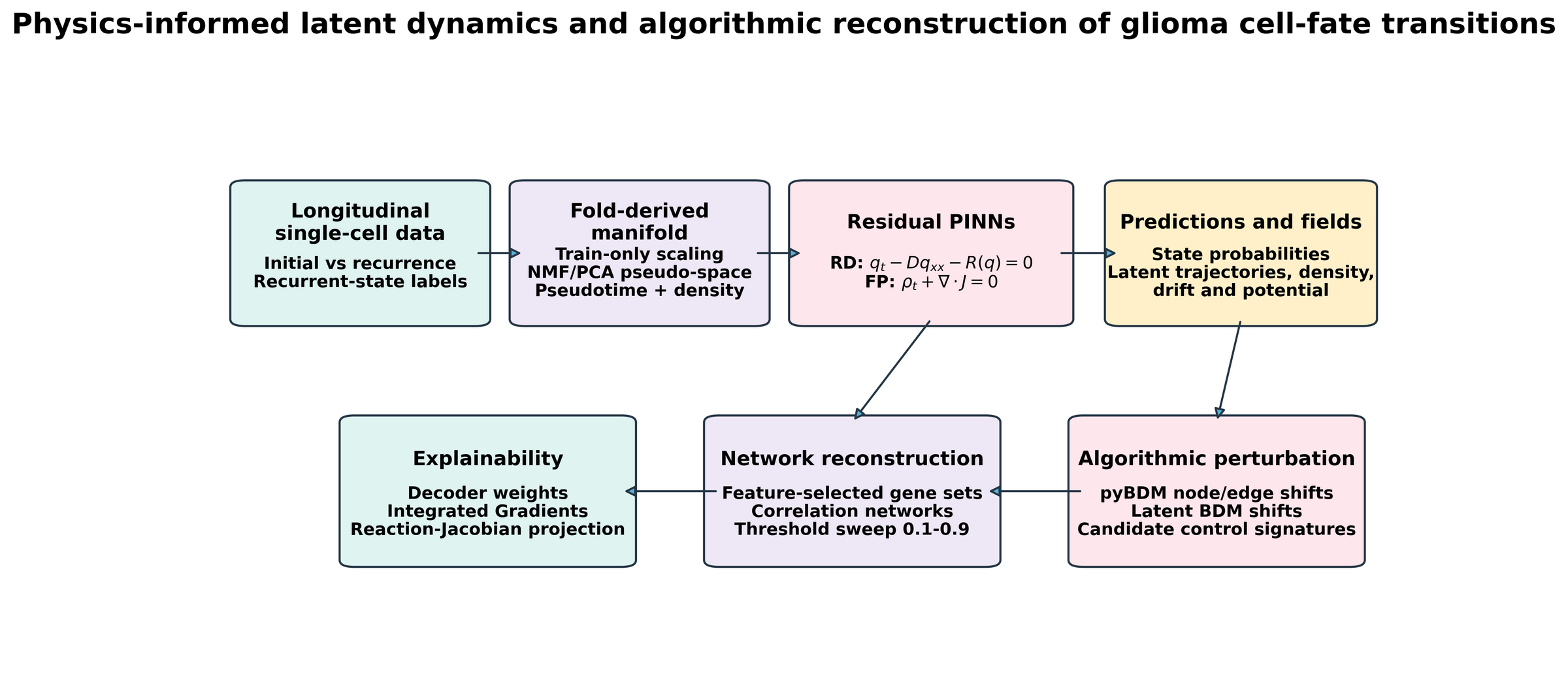}
\caption{Study design. Longitudinal single-cell data were mapped to fold-specific pseudo-space and pseudotime, modeled by canonical residual-based PINNs, decoded for XAI, and analyzed with gene-network and latent-space BDM perturbations. Only the RD and FP coordinate-residual models are designated canonical PINNs; the 11-model roster is treated as a set of physics-structured or dynamical systems comparators.}
\label{fig:workflow}
\end{figure}

\section{Methods}
\subsection{Data objects, labels and cross-validation}
Primary biological interpretation was restricted to H3K27M patient data from Sussman er al. with informative binary endpoints: (i) initial versus recurrent cells from the longitudinal object; and (ii) an annotation-derived recurrent-state axis contrasting stem/progenitor-like with differentiated/glial-like cells \citep{sussman2026atlas}. The source metadata retained multiple fields---including major cell type, Neftel classification cell type, identity, merged cell type and updated identity---but this simple study did not  implement a validated multiclass cell-type classification problem. These fields were integrated into the prespecified binary recurrent-state target in the neural models and used for stratified sampling to ensure accurate representation of all cellular subtypes/phenotypes within the profiled tumor ecology. We therefore interpret the second task as a coarse state-axis model, not as a definitive lineage taxonomy or a stem-versus-all-other-cells classifier. 

Cells were capped at 20,000 by stratified sampling for computational purposes with architectures trained and evaluated using a five-fold cross-validation. Scaling, non-negative matrix factorization (NMF), coordinate construction and all trainable transformations were fit within each training fold. NMF factorized a non-negative expression matrix $V$ into $W H$, yielding interpretable metagene coordinates and programs \citep{lee1999learning}. Pseudo-space, pseudotime and density were derived from the fold-specific representation. These coordinates organize the observed cell fate manifold but are not assumed to be literal anatomical coordinates or prospectively observed biological time.

\subsection{Shared representation and optimization}
Gene expression was encoded through a multilayer autoencoder to an eight-dimensional latent state to reflect the number of cellular subtypes (phenotypic subgroups) in glioma ecology population dynamics. The decoder mapped the latent state back to genes, and an $8\rightarrow64\rightarrow C$ classifier predicted cell state. The final representation used for decoding and classification was normally
\begin{equation}
z_{\mathrm{fused}}=0.9\,z_{\mathrm{PDE}}+0.1\,z_{\mathrm{enc}},
\end{equation}
where the encoder contribution stabilizes reconstruction while retaining a predominantly PDE-constrained state. Coordinate multilayer perceptrons used hyperbolic tangent activations, Xavier initialization and min--max scaling to $[-1,1]$. Each fold used 1,500 Adam steps followed by up to 250 strong-Wolfe L-BFGS iterations, with 4,096 interior collocation points, 256 initial-condition points and 256 boundary points on an NVIDIA T4 GPU. Hyperparameter selection was primarily guided by the structure and dimensionality of the data, together with metadata-informed expectations regarding the number of biologically meaningful cellular states.

\subsection{Reaction--diffusion PINN}
For eight latent fields $q=(q_1,\ldots,q_8)$, separated into activator- and inhibitor-like channels, the RD model enforced
\begin{equation}
\mathcal{R}_{\rd}(x,t)=\partial_t q-D\,\partial_{xx}q-R_\theta(q)=0,
\end{equation}
where $D$ contains learned positive diffusivities and $R_\theta:\RR^8\rightarrow\RR^8$ is an $8\rightarrow64\rightarrow64\rightarrow8$ reaction network. The coordinate field used eight 64-unit tanh layers. Early-pseudotime cells supplied initial anchors and $\partial_x q=0$ supplied no-flux boundaries. PDE, initial and boundary weights were self-adaptive and bounded as $\lambda=0.1+9.9\,\sigma(a)$. The complete objective combined data anchoring, PDE residual, boundary and initial losses, reconstruction and classification.

\subsection{Fokker--Planck PINN}
The FP model mapped $(x_1,x_2,t)$ through seven 64-unit tanh layers to eight latent fields and a positive density $\rho$. A separate potential network generated drift $b=-\nabla V_\phi$. With flux $J=\rho b-D_\rho\nabla\rho$, probability conservation required
\begin{equation}
\mathcal{R}_{\rho}=\partial_t\rho+\nabla\cdot J=0.
\end{equation}
Latent fields simultaneously obeyed
\begin{equation}
\mathcal{R}_{q}=\partial_t q+b\cdot\nabla q-\kappa\Delta q=0.
\end{equation}
The objective included both residuals, a no-normal-flux boundary condition, early density and state anchors, experimental latent anchors, reconstruction and classification. Positive $D_\rho$ and $\kappa$ were learned by softplus parameterization.

\begin{figure}[H]
\centering\includegraphics[width=\linewidth,height=0.78\textheight,keepaspectratio]{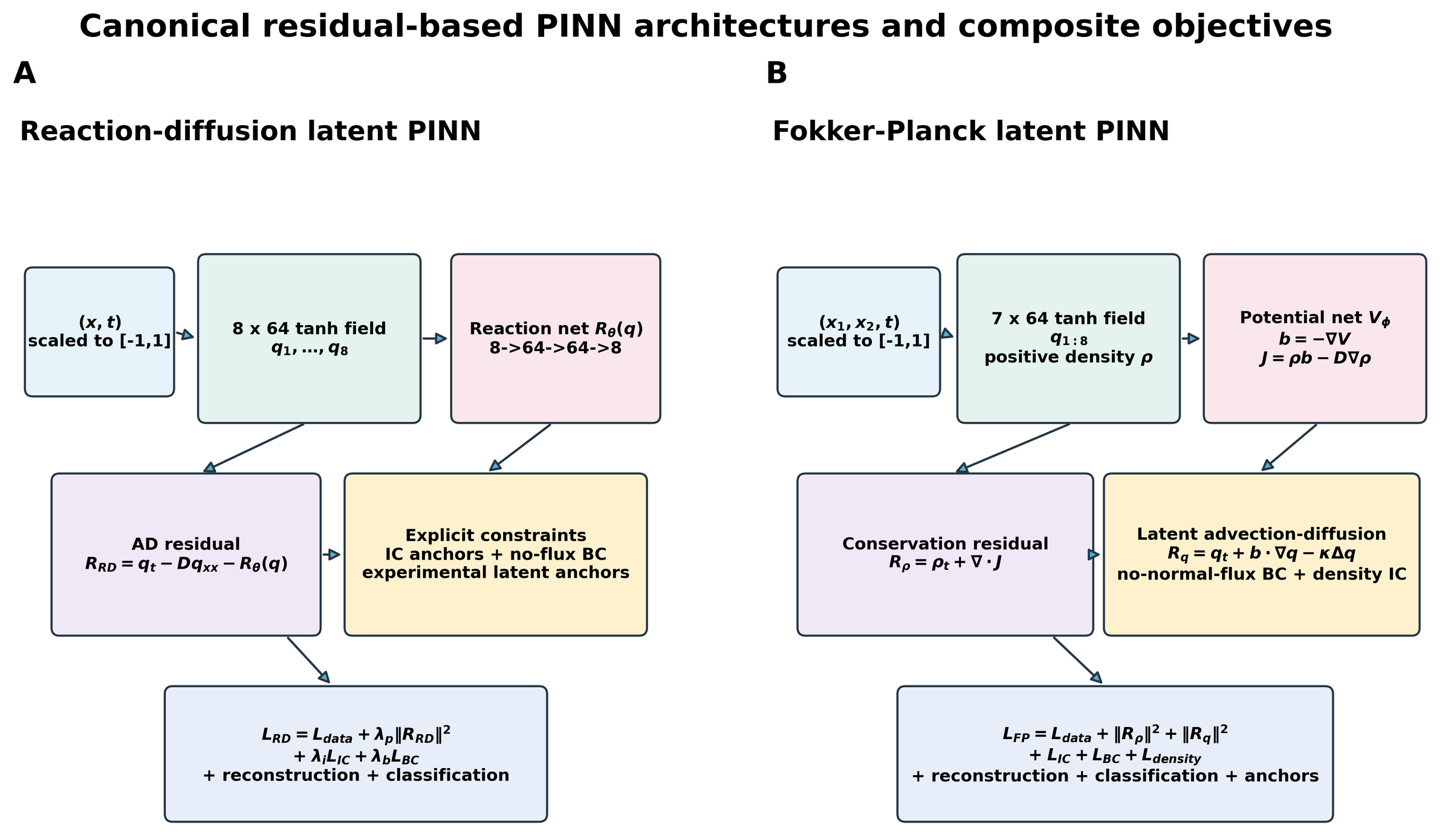}
\caption{Canonical residual-based architectures. Both PINNs optimize automatic-differentiation equation residuals at interior collocation points plus explicit initial, boundary, data, reconstruction and classification losses. Architectural and complete loss details are provided in the Supplement.}
\label{fig:architectures}
\end{figure}

\subsection{XAI, landscapes and algorithmic perturbations}
Gene importance was measured by decoder-column norm, Integrated Gradients for reconstruction, classification, encoder--PINN alignment and total latent activity, RD activator/inhibitor attribution, and a decoder-propagated reaction Jacobian. Waddington-like energy surfaces were estimated from the empirical density as $U(z)=-\log[P(z)+\epsilon]$ and used as descriptive landscape visualizations rather than direct measurements of developmental potential.

For each feature set, gene--gene association matrices were thresholded at $0.1,0.3,0.5,0.7$ and $0.9$. BDM was computed for binarized one- and two-dimensional objects. For a decomposition into distinct blocks $r_u$ with multiplicities $n_u$,
\begin{equation}
\bdm(X)=\sum_u \left[\ctm(r_u)+\log_2 n_u\right].
\end{equation}
Node perturbation was defined exactly as
\begin{equation}
\Delta\bdm(v)=\bdm(G\setminus v)-\bdm(G),
\end{equation}
with analogous edge deletion. Rankings in the main figures use mean absolute shifts across available model--XAI--threshold instances to emphasize perturbation magnitude; signed threshold-specific shifts are retained. Shannon entropy, entropy-rate estimates and lossless compression gzip, zlib and LZMA shifts were calculated as comparators. Learned latent trajectories were ordered by pseudotime, binarized at quantiles $0.25$, $0.50$ and $0.75$, and analyzed for class, latent-dimension and latent-network BDM shifts.

\subsection{Evaluation}
Classification was evaluated by accuracy, macro-F1 and area under the receiver-operating-characteristic curve (AUC). Reconstruction was evaluated in standardized expression space by mean squared error (MSE), mean absolute error (MAE) and $R^2$. Means and sample standard deviations were calculated across five folds. The selected-feature longitudinal expression analysis used Mann--Whitney tests with Benjamini--Hochberg correction and effect sizes after appropriate normality and residual variance tests. Since features were selected through the modeling workflow, this analysis is descriptive and not an independent genome-wide discovery test.

\subsection{Focused gene-set enrichment}
For downstream gene set enrichment and pathway analysis, RD-PINN was selected because it outperformed FP-PINN on both primary tasks. This selection does not imply superiority across the broader comparator roster or transfer to other datasets or disease ecosystems. For each task, we selected the top 30 genes by five-fold mean absolute class-IG importance and, separately, by mean absolute node-perturbation $\Delta\bdm$ across the five thresholds of the RD union-feature network. No selection-boundary ties occurred. This BDM analysis concerns one model's union signature network, distinct from the pooled rankings in Fig.~\ref{fig:bdm}. Symbols were resolved to HGNC-approved names; unresolved or ambiguous mappings were excluded without replacement, and mitochondrial/ribosomal genes were retained. Python/SciPy one-sided hypergeometric tests used frozen Enrichr GO Biological Process 2023, Reactome 2022 and KEGG 2021 Human libraries \citep{enrichr2016,go2023,reactome2022,kegg2021}. Benjamini--Hochberg correction covered all 30,180 tests (four signatures by 7,545 terms), including zero overlaps \citep{bh1995}. The 16,614-gene annotation-union background was exploratory because the full measured-gene universe was unavailable. A separate BDM sensitivity analysis used eligible union-network genes as background and corrected 2,573 tests jointly. 

\section{Results}
\subsection{Residual PINNs separated longitudinal and recurrent-state labels}
\begin{figure}[H]
\centering\includegraphics[width=\linewidth,height=0.78\textheight,keepaspectratio]{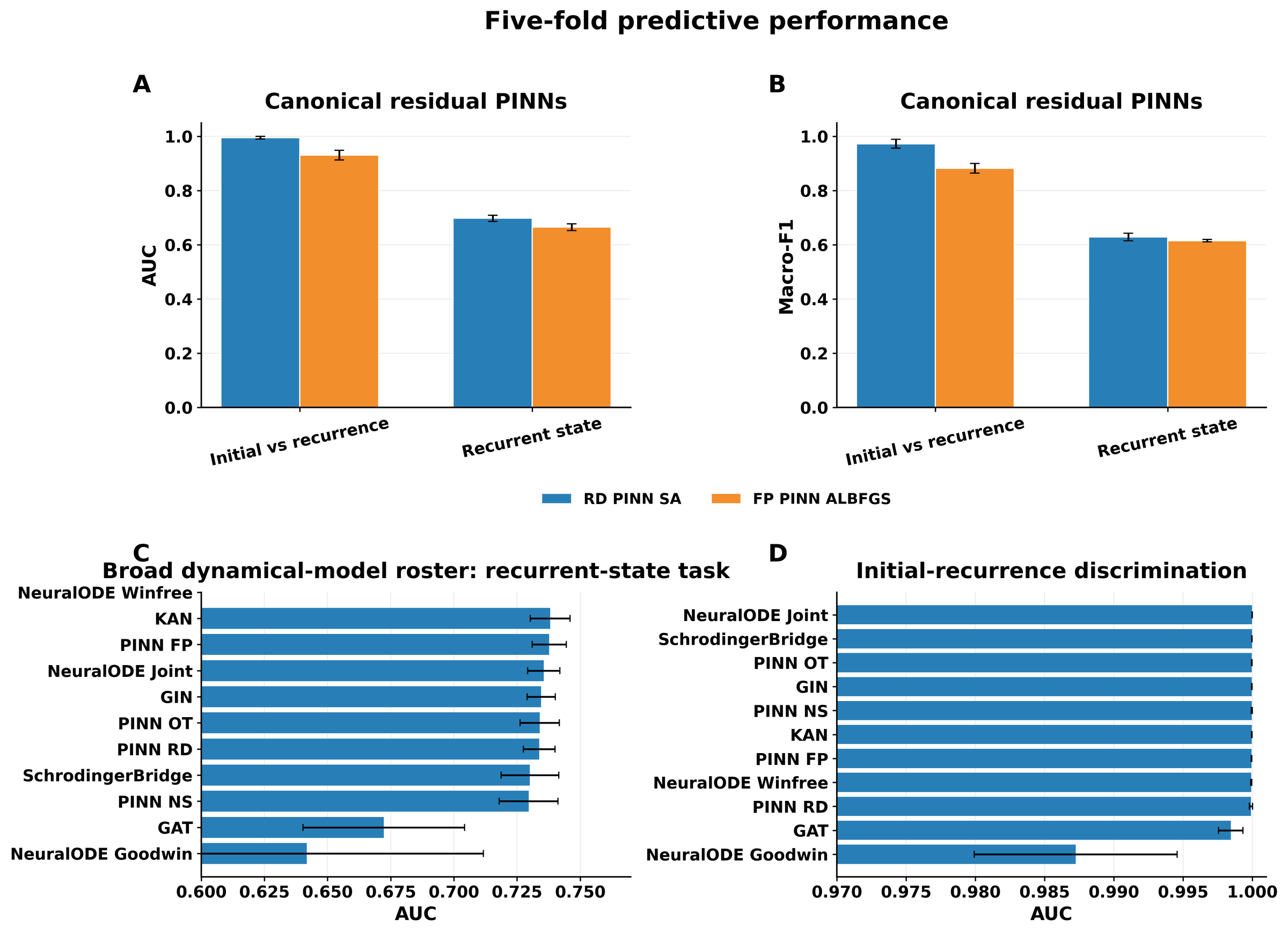}
\caption{Predictive performance. (A--B) Five-fold AUC and macro-F1 for the two residual PINNs. (C) Recurrent-state AUC across the broader dynamical systems roster. (D) Near-ceiling initial--recurrence discrimination across broad comparators, while training confounds and data leakage were prevented.}
\label{fig:performance}
\end{figure}

The RD model showed stronger discrimination than FP in both primary tasks (Table~\ref{tab:performance}; Fig.~\ref{fig:performance}). For initial versus recurrence, RD achieved AUC $0.995\pm0.005$, accuracy $0.979\pm0.013$ and macro-F1 $0.973\pm0.016$. FP achieved AUC $0.931\pm0.018$, accuracy $0.908\pm0.015$ and macro-F1 $0.882\pm0.018$. Within recurrence, where stem-like and differentiated-like annotations were less cleanly separable, RD achieved AUC $0.698\pm0.012$ and macro-F1 $0.629\pm0.014$, compared with FP AUC $0.665\pm0.012$ and macro-F1 $0.615\pm0.004$.

\begin{table}[H]
\centering\small
\caption{Five-fold performance of the two canonical residual-based PINNs. Values are mean $\pm$ SD.}
\label{tab:performance}
\begin{tabular}{llccc}
\toprule
Task & Model & AUC & Accuracy & Macro-F1\\
\midrule
Initial vs recurrence & RD-PINN & $0.995\pm0.005$ & $0.979\pm0.013$ & $0.973\pm0.016$\\
Initial vs recurrence & FP-PINN & $0.931\pm0.018$ & $0.908\pm0.015$ & $0.882\pm0.018$\\
Recurrent state & RD-PINN & $0.698\pm0.012$ & $0.773\pm0.016$ & $0.629\pm0.014$\\
Recurrent state & FP-PINN & $0.665\pm0.012$ & $0.723\pm0.012$ & $0.615\pm0.004$\\
\bottomrule
\end{tabular}
\end{table}

The broader 11-model roster produced recurrent-state AUCs from approximately 0.64 to 0.74 among models with defined AUC, with KAN, the physics-structured FP comparator, the joint Neural ODE and GIN among the upper group. In contrast, many broad models reached near-ceiling initial--recurrence AUC. At this pilot stage, this should not be interpreted as resolved forecasting: longitudinal labels can be entangled with patient, sampling time, treatment and selected-feature structure, all of which were accounted for and removed from the dataset training. However, these differences can organize the learnable latent space and cellular pattern distributions. While the initial--recurrence task showed reproducible and robust discrimination, the more difficult recurrent-state task of within-recurrence organization produced modest performance metrics. This demonstrates the level of complexity among the diverse subtypes and phenotypic plasticity spectrum within-tumor state ecology.

\subsection{PDE-constrained representations formed structured energy landscapes}
\begin{figure}[H]
\centering\includegraphics[width=\linewidth,height=0.78\textheight,keepaspectratio]{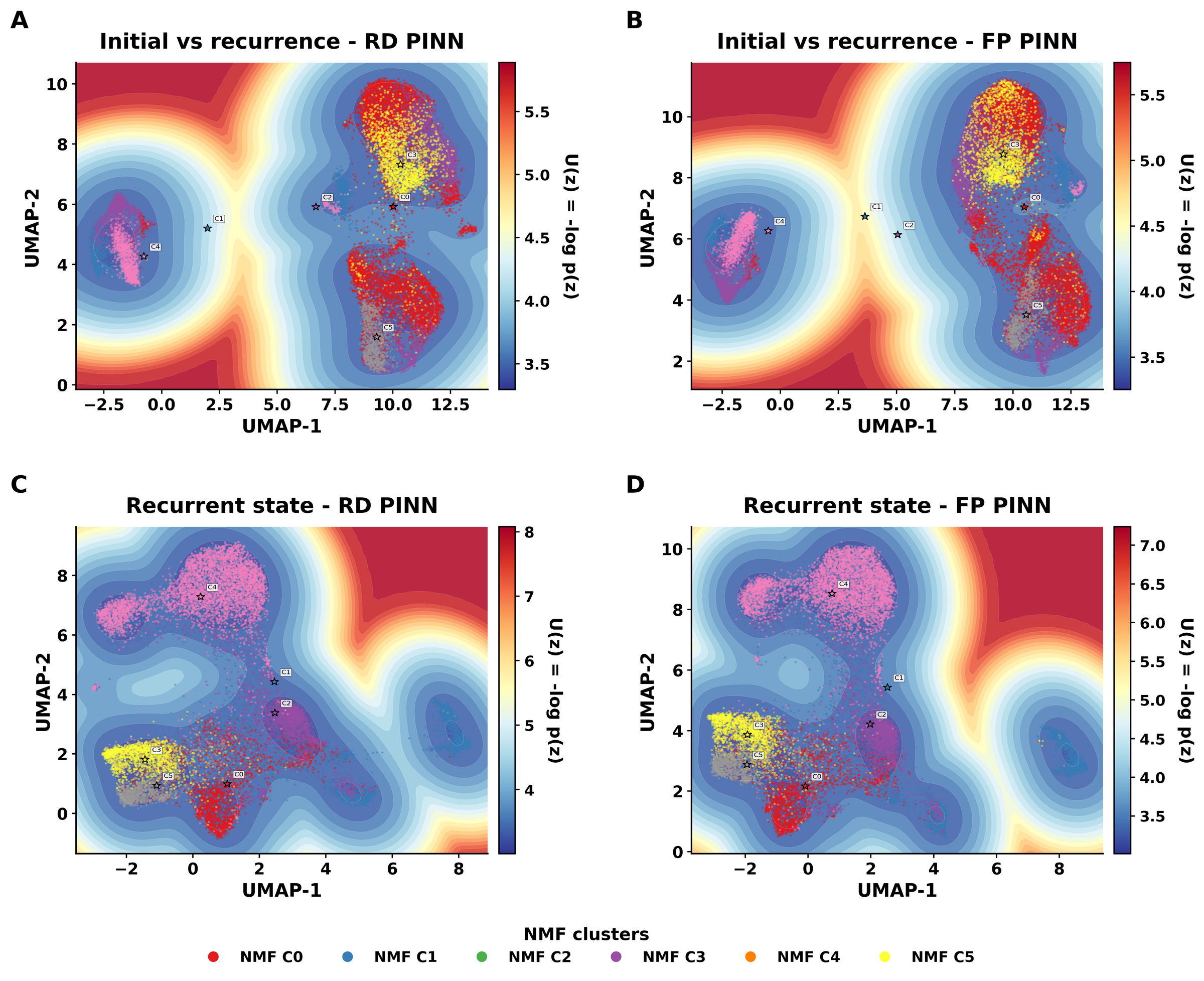}
\caption{Waddington-like energy landscapes reconstructed from feature-informed latent representations. Energy is $-\log$ empirical density. Stars denote identified basin centers. The surfaces are descriptive manifold learning summaries.}
\label{fig:landscapes}
\end{figure}

Density-derived energy surfaces contained multiple basins and transition regions in both tasks (Fig.~\ref{fig:landscapes}). RD and FP generated broadly concordant large-scale geometry while differing in local basin shape. The recurrent-state landscapes were less profoundly separated than the initial--recurrence landscapes, consistent with the lower recurrent-state AUC. These visualizations support heterogeneity and manifold structure and infer plausible biological attractors. However, these dynamical features can only be established with longitudinal lineage tracing and experimental perturbation assays.

\subsection{Cross-model XAI recovered reproducible transition-associated genes}
\begin{figure}[H]
\centering\includegraphics[width=\linewidth,height=0.78\textheight,keepaspectratio]{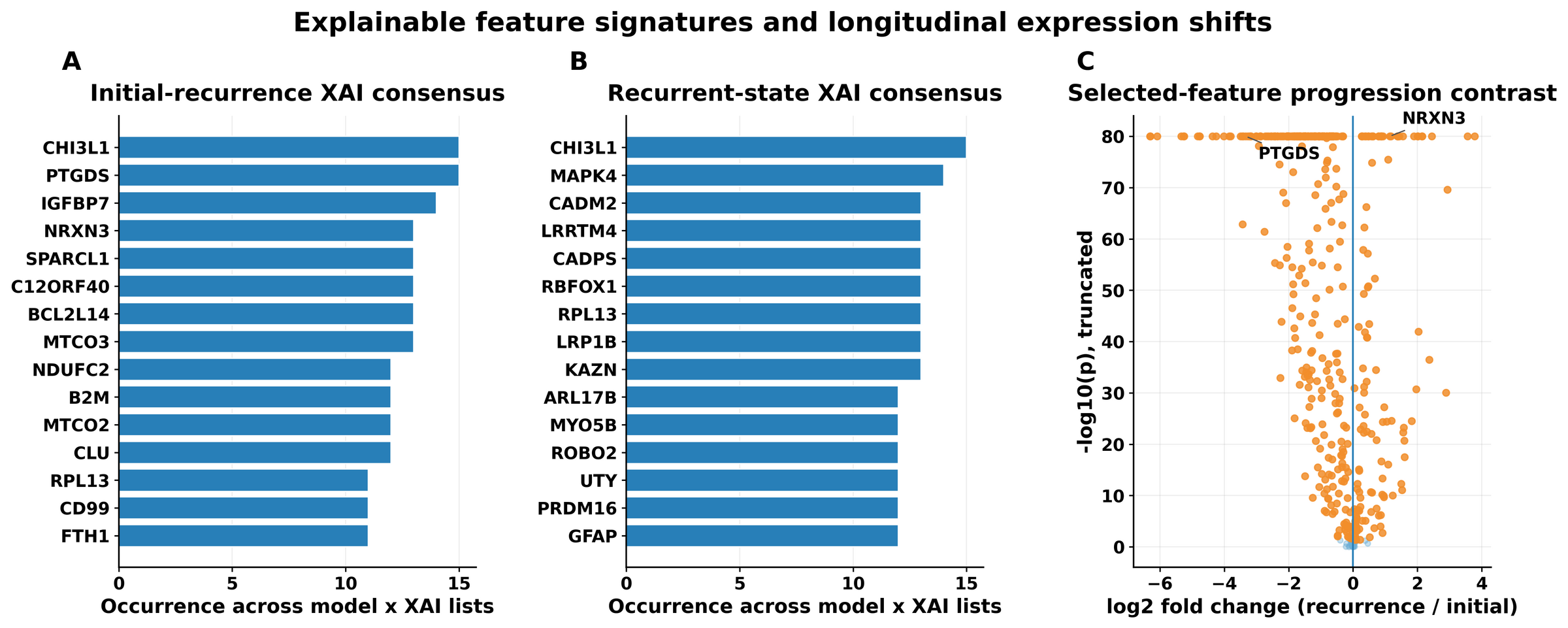}
\caption{Explainable signatures. (A--B) Frequency of top genes across model and attribution outputs. (C) Longitudinal expression contrast restricted to the modeling-selected feature union; the plotted significance axis is truncated because several rank-test probabilities underflowed numerically.}
\label{fig:xai}
\end{figure}

Cross-model recurrence-frequency rankings integrated decoder, Integrated Gradient and dynamical systems-Jacobian feature lists (Fig.~\ref{fig:xai}). In initial--recurrence analysis, \textit{PTGDS} and \textit{CHI3L1} appeared in 15 lists, \textit{IGFBP7} in 14, and \textit{SPARCL1}, \textit{BCL2L14}, \textit{C12ORF40}, \textit{NRXN3} and mitochondrial features in 13. In recurrent-state analysis, \textit{CHI3L1} appeared in 15 lists, \textit{MAPK4} in 14, and \textit{LRP1B}, \textit{KAZN}, \textit{RPL13}, \textit{RBFOX1}, \textit{LRRTM4}, \textit{CADPS} and \textit{CADM2} in 13.

The descriptive selected-feature progression contrast was directionally heterogeneous. \textit{CHI3L1}, \textit{SPARCL1}, \textit{C12ORF40}, \textit{PTGDS}, \textit{IGFBP7}, \textit{BCL2L14}, \textit{FTH1}, \textit{B2M} and \textit{CD63} were lower in recurrence, whereas \textit{NRXN3}, \textit{NRG1}, \textit{LRRTM4} and \textit{MMP16} were higher. For example, log$_2$ recurrence/initial fold changes were $-2.89$ for \textit{CHI3L1}, $-3.33$ for \textit{PTGDS}, $-2.48$ for \textit{IGFBP7}, $+1.14$ for \textit{NRXN3} and $+1.18$ for \textit{NRG1}. These shifts describe the analyzed cells and selected genes; they do not isolate treatment effects from cohort composition.

\subsection{BDM perturbations identified network-sensitive stress and immune nodes}
\begin{figure}[H]
\centering\includegraphics[width=\linewidth,height=0.78\textheight,keepaspectratio]{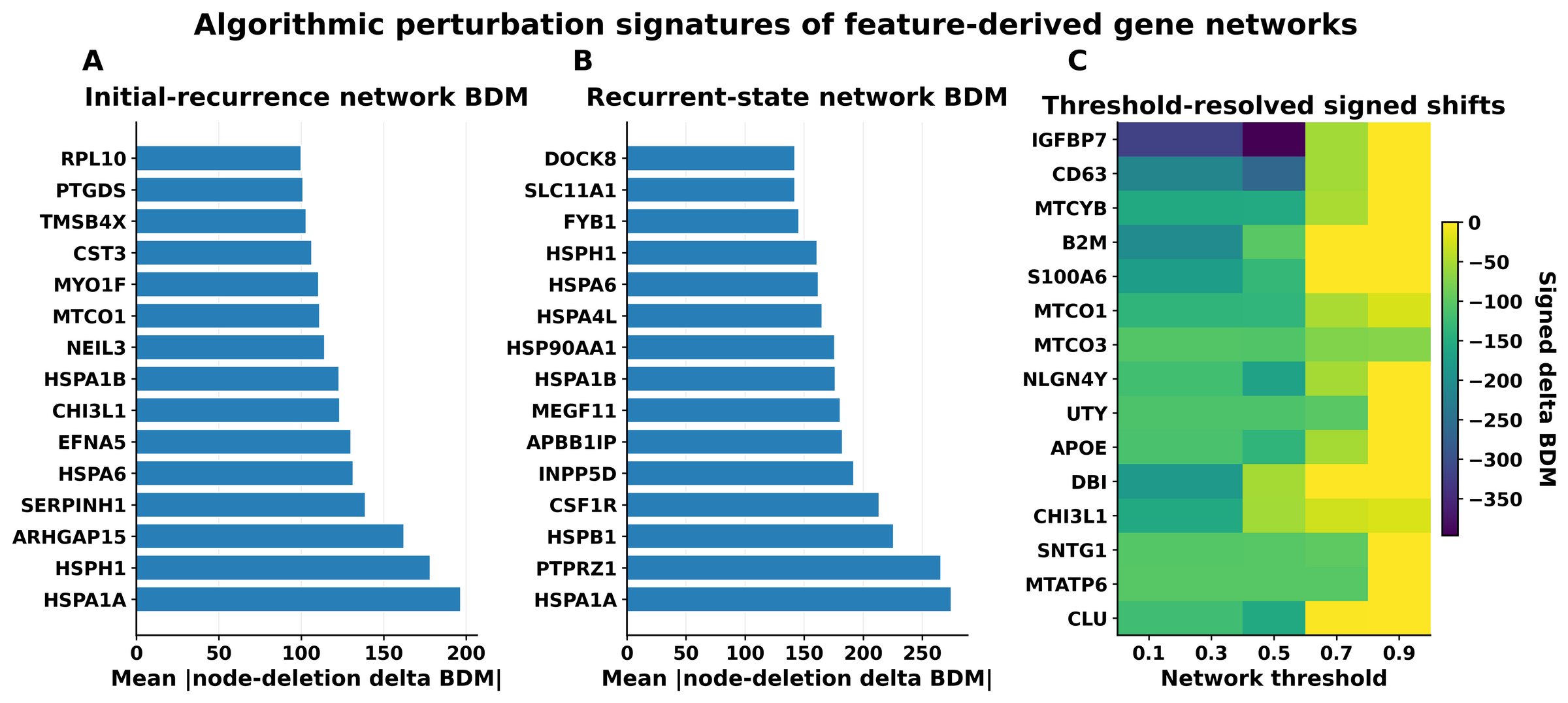}
\caption{Gene-network algorithmic perturbations. (A--B) Mean absolute node-deletion $\Delta\bdm$ pooled across available canonical-PINN model, XAI-channel and threshold instances. (C) Signed RD union-network shifts across thresholds, demonstrating scale sensitivity.}
\label{fig:bdm}
\end{figure}

BDM rankings were not identical to but complemented XAI rankings (Fig.~\ref{fig:bdm}). XAI asks which inputs influence the learning of a model output, while edge or node-deletion $\Delta\bdm$ asks how much the approximate minimal description length of a thresholded complex network changes when that network feature is removed. Across available canonical-PINN feature channels and thresholds, high initial--recurrence perturbation magnitudes included \textit{HSPA1A}, \textit{HSPH1}, \textit{HSPA6}, \textit{CHI3L1}, \textit{HSPA1B}, \textit{MTCO1}, \textit{CST3}, \textit{PTGDS}, \textit{CD63} and \textit{IGFBP7}. Within recurrent-state networks, prominent nodes included \textit{HSPA1A}, \textit{HSPB1}, \textit{INPP5D}, \textit{APBB1IP}, \textit{HSPA1B}, \textit{HSP90AA1}, \textit{HSPA4L}, \textit{HSPA6}, \textit{HSPH1}, \textit{FYB1}, \textit{SLC11A1} and \textit{DOCK8}. Threshold-resolved shifts remained strongly signed and scale dependent, motivating reporting of both signed matrices and absolute consensus magnitudes.

\subsection{Algorithmic perturbation shifts were also detectable directly in learned latent states}
\begin{figure}[H]
\centering\includegraphics[width=\linewidth,height=0.78\textheight,keepaspectratio]{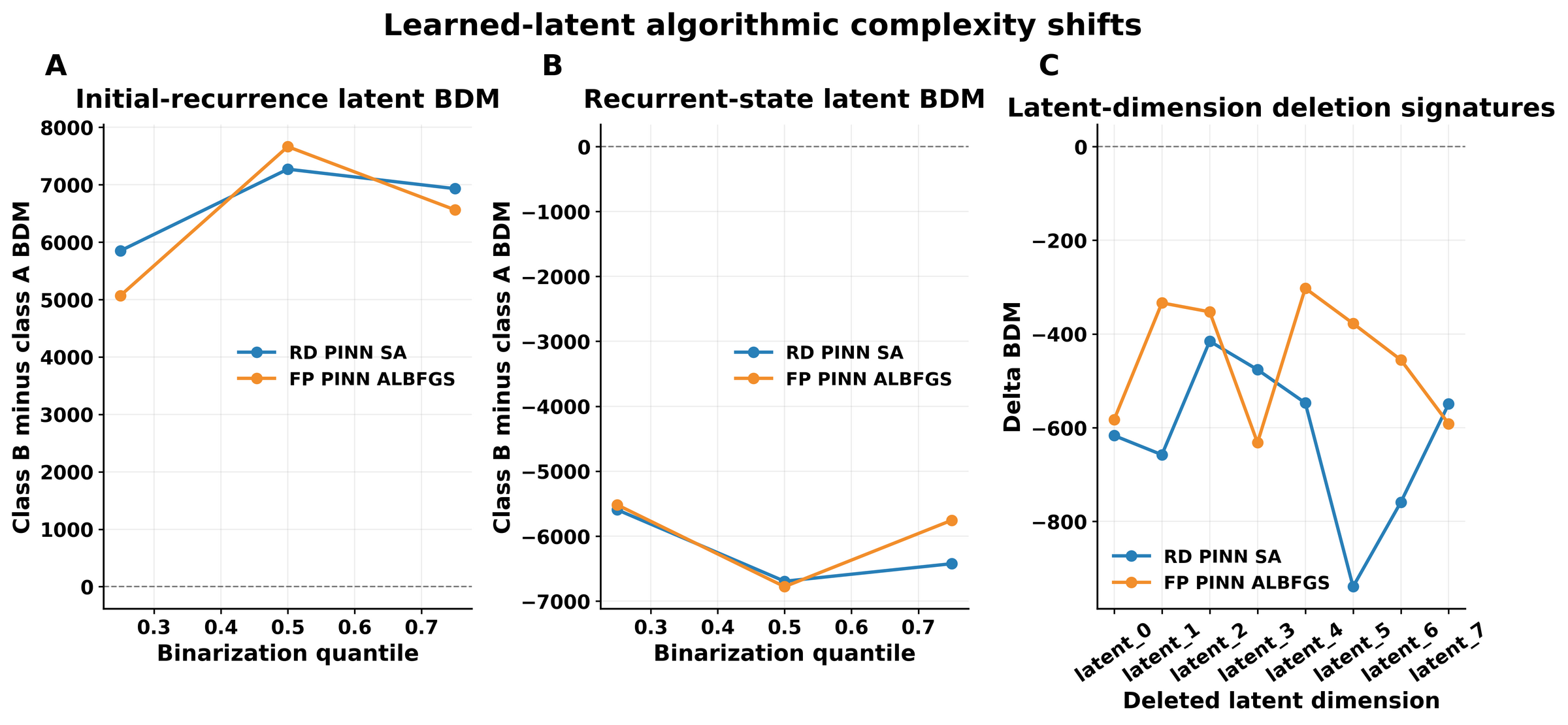}
\caption{Learned-latent BDM analysis. (A--B) Class-to-class BDM differences across binarization quantiles. (C) BDM change after deleting each latent dimension from the ordered latent matrix. Sign depends on the exported label order and perturbation definition; magnitudes indicate representation sensitivity rather than biological causality.}
\label{fig:latentbdm}
\end{figure}

The posthoc workflow measured BDM on pseudotime-ordered, binarized latent matrices rather than only on decoded gene networks (Fig.~\ref{fig:latentbdm}). In initial--recurrence RD representations, recurrence-minus-initial BDM shifts were positive at all three binarization quantiles (approximately 5,849, 7,269 and 6,929). FP showed the same direction with model-dependent magnitude. Recurrent-state class shifts were negative in the exported label order, underscoring that direction depends on class coding and representation. Deleting individual latent dimensions reduced matrix BDM in the initial--recurrence RD representation, with latent dimension 5 producing the largest magnitude in that analysis ($\Delta\bdm\approx-939$). Thus the learned manifold geometry contains nonuniform algorithmic contributions that are not captured by a single scalar complexity score.

\subsection{RD-PINN signature enrichment distinguishes progression and recurrent-state annotations}
\begin{figure}[H]
\centering\includegraphics[width=\linewidth,height=0.70\textheight,keepaspectratio]{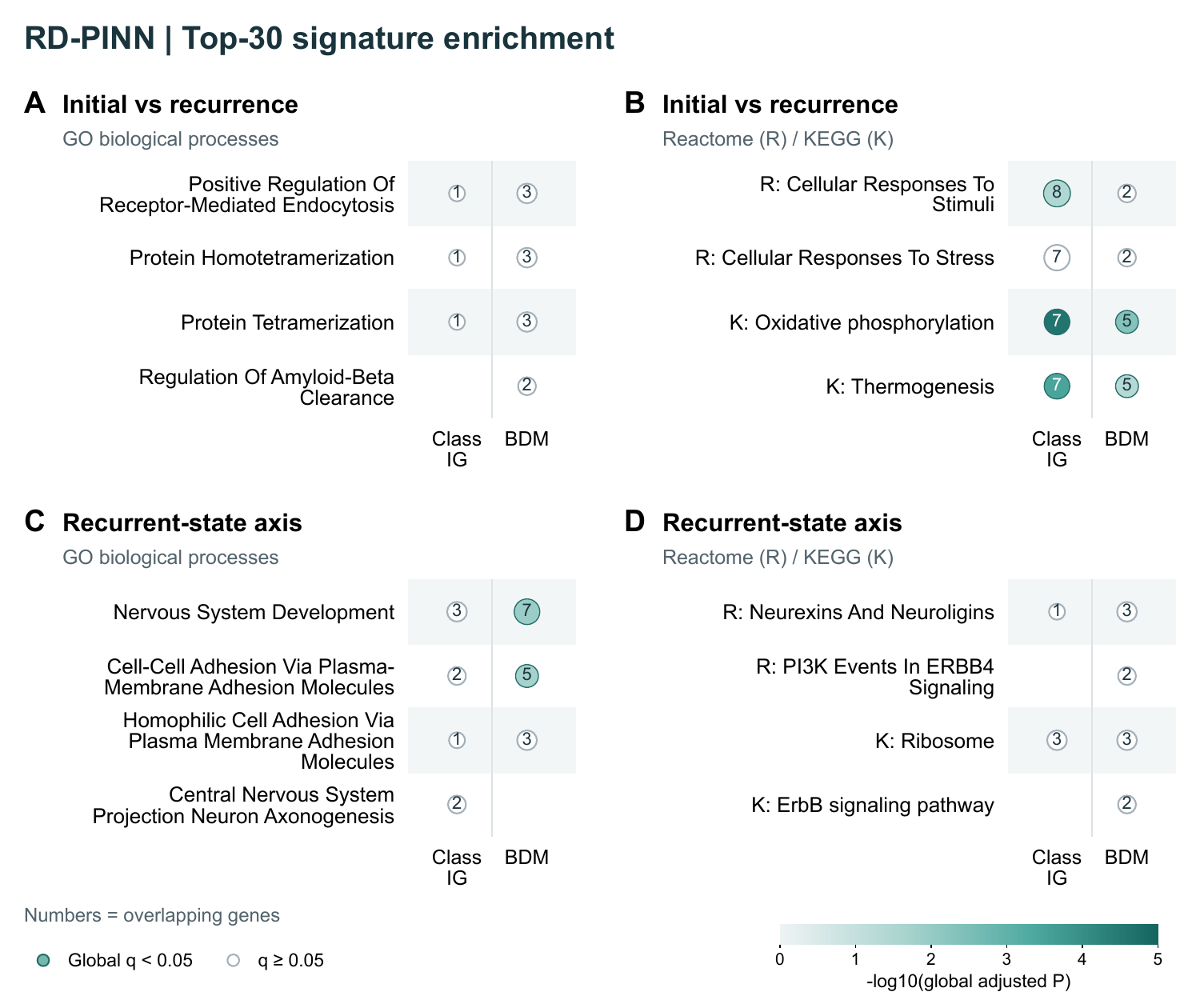}
\caption{\textbf{Top-30 RD-PINN signature enrichment.} (A--B) Initial versus recurrence; (C--D) the recurrent-state axis. Columns compare class integrated gradients (IG) with BDM from the RD union-feature network. Circle size and number encode gene overlap; filled circles indicate global $q<0.05$, colour encodes $-\log_{10}q$, open circles indicate nonsignificant overlap, and blanks indicate zero overlap. Each task displays four GO terms and two terms each from Reactome (R) and KEGG (K), ranked by the minimum global $q$ across the two signatures, with raw $P$ and name resolving ties and at least two overlapping genes in one signature.}
\label{fig:rd-enrichment}
\end{figure}

The focused analysis identified 18 significant signature--term combinations, including overlapping annotations (Fig.~\ref{fig:rd-enrichment}). Progression class-IG and BDM signatures both enriched oxidative phosphorylation ($q=2.38\times10^{-5}$ and $0.00472$; seven and five overlapping genes) and thermogenesis ($q=0.000344$ and $0.0415$). Recurrent-state BDM enriched nervous-system development ($q=0.0121$; seven genes) and cell--cell adhesion via plasma-membrane adhesion molecules ($q=0.0189$; five genes, including \textit{CADM2}, \textit{LRRC4C} and \textit{ROBO2}); no recurrent-state class-IG term met global $q<0.05$. No BDM term remained significant against its eligible network background. Thus, the results summarize annotations of selected features, including mitochondrial contributions, and do not establish pathway activation or enrichment independent of upstream feature selection.

\section{Discussion}
This study integrates three levels of glioma ecology description. Residual PINNs impose explicit dynamical hypotheses on a single-cell fate geometry (manifold). XAI maps the learned fields and classifiers back to predictive signatures (genes). BDM perturbation analysis asks which molecular or latent components most alter estimated algorithmic description length of the network. The complexity-informed combination is useful precisely because the multi-layers answer distinct questions in the study of glioma cell fate decisions as complex adaptive systems. Predictive accuracy measures separability, PDE residuals encode consistency with a chosen dynamical form, XAI measures model sensitivity, and $\Delta\bdm$ measures perturbation-sensitive self-organization of a discretized object. Agreement strengthens prioritization, while disagreement can reveal different aspects of the complex system's representation learning.

The RD model consistently outperformed FP in the two primary tasks, although it still struggled with the more complex within-recurrence task. Several interpretations are possible. A learned local reaction field with channel-specific diffusion may be a better inductive bias for the constructed pseudo-morphospace. The one-dimensional RD coordinate may be easier to optimize than joint density conservation and potential-driven transport, or the FP objective may distribute capacity across nonequilibrium statistical properties such as density, drift and latent-state equations. These findings do not imply that tissue evolution or glioma ecology is literally governed by the fitted RD equation. Rather, they show that a reaction--diffusion residual was an effective and inspectable structural constraint for the observed latent manifold geometry.

The RD--FP performance difference also warrants an optimization interpretation and hyperparameter tuning in future experiments. Differential-operator penalties can make PINN objectives difficult to optimize, so predictive differences need not identify the most faithful biological equation \citep{krishnapriyan2021failure}. The present models differ in coordinate dimension, field parameterization and constrained quantities. Their comparison therefore evaluates complete modeling choices rather than isolating the PDE term alone. Matched-coordinate and matched-capacity Neural ODE/PINN ablations, together with held-out residual and condition errors, would help separate useful dynamical constraints from architectural and training effects.

Regardless, our pilot study demonstrates that XAI and BDM results point to complementary biological programs steering cell fate geometry. Longitudinal XAI emphasized secreted, extracellular, glial, stress and neuronal-interaction features, including \textit{PTGDS}, \textit{CHI3L1}, \textit{IGFBP7}, \textit{SPARCL1}, \textit{NRXN3} and \textit{NRG1}. Recurrent-state XAI added genes related to signaling, adhesion/matrisome and neural identity. Both converged on several developmental and neuronal fate regulators. BDM perturbations shifted toward heat-shock and immune-network organization. This difference is expected: a feature can strongly influence classification without occupying a globally organization-sensitive network position, and a network node can strongly alter description length without being the largest local gradient feature. Candidate network signatures supported by both lenses, such as \textit{CHI3L1}, \textit{PTGDS}, \textit{IGFBP7} and stress-response genes, are priorities for targeted validation.

The near-ceiling initial--recurrence discrimination in the broad roster illustrates the learnability of discriminant analysis. It indicates that the analyzed cellular states contain a strong longitudinal separation, but that separation may incorporate patient identity, treatment exposure, specimen processing, cell composition and annotation structure that could have influenced these results. While we prevented their leakage into the data training and validation, the time batches could be inherent to the pattern discovery and learnable novelty of data structures. The recurrent-state task, with modest AUC values near 0.70, is more consistent with the expected difficulty of plastic cell-state classification. 

Therefore, future work should use patient-grouped or leave-patient-out validation wherever sample counts permit, preserve treatment and spatial covariates, and test signatures in external cohorts and perturbation experiments. The sparse and limited dataset prevented such validation in the current pilot. 

From a complex systems perspective, our deep learning and AID merged framework treats glioma not as a snapshot of markers but as a constrained, interacting, multiscale dynamical process; a complex adaptive ecosystem. ``More is different'' because glioma ecology includes nonlinear relationships and adaptive feedback loops embedded within tumor-immune-multisystemic environments, beyond the sum of its parts\citep{anderson1972more}. Our goal herein was to repurpose causal inference tools from computational and systems medicine to help identify reproducible state-transition structure, early-warning features, and candidate intervention points that could eventually inform risk stratification or cell fate reprogramming candidates. Moving from computational signatures to precision oncology will require prospective sampling, lineage information, spatial validation, mechanistic experiments (such as CRISPR screens, RNA interference and pharmacological perturbations) and clinically meaningful endpoints.

\section{Limitations}
The principal limitation is the use of NMF/PCA-derived pseudo-space and pseudotime. The canonical PINNs satisfy the numerical residual-based definition---they minimize AD-based PDE residuals with explicit conditions---but the coordinates are learned organizational variables, not literal anatomical space and clock time. Second, initial--recurrence prediction can be confounded by patient and timepoint structure even though they were prevented/removed in the model training pipeline. Third, the recurrent-state endpoint was derived from multiple annotation fields and therefore inherits annotation uncertainty and coarse binary grouping. No validated multiclass cell-type claim is made herewith. Fourth, BDM values depend on binarization, network thresholds, block decomposition and gene-set construction.  Finally, computational perturbation is candidate causal discovery (hypotheses generation) not a substitute for genetic, pharmacological or lineage-tracing intervention. Multi-cohort and multi-institutional validation patient datasets will be required in future translation of our models. 

\section{Conclusion}
In this study, we used physics-informed artificial intelligence and complex systems to study pHGG dynamics, and predict cell fate trajectories. Canonical residual-based latent PINNs, XAI and BDM perturbation analysis provide a unified but carefully qualified framework for studying glioma cell-fate dynamics. The RD model gave the strongest five-fold performance. XAI identified reproducible longitudinal and recurrent-state genes; and network and latent $\Delta\bdm$ exposed intervention-sensitive network self-organization not reducible to some scalar quantity or single predictive score. The complex systems framework is hereby positioned as an in silico causal inference tool for attractor-oriented reconstruction and candidate mechanism prioritization, with experimental validation required before biomarker or therapeutic interpretation.

\section*{Code and data availability}
The analysis was executed with CUDA/T4 pipeline. The canonical PINN implementation was adapted at the architectural-pattern level from two public repositories \citep{bhustaliPINNrepo,kimPINNrepo}. All glioma-specific data loading, latent manifolds, PDE systems, XAI and BDM posthoc analyses were implemented in our original codes and analyzed pipeline. All original code required to reproduce our findings is provided at \href{https://github.com/Abicumaran/Single-cell-PINNs-and-NeuralODEs}{https://github.com/Abicumaran/Single-cell-PINNs-and-NeuralODEs}. The repository will be made public post-publication of the paper.

\section*{Competing Interests}
The author declares there are no competing interests.

\bibliographystyle{plainnat}
\bibliography{references}
\end{document}